# Construction of a Schwarzschild-Couder telescope as a candidate for the Cherenkov Telescope Array: status of the optical system

**J. Rousselle**[1], **K. Byrum**[2], **R. Cameron**[3], **V. Connaughton**[4], **M. Errando**[5], **V. Guarino**[2], **T. B. Humensky**[6], **P. Jenke**[4], **D. Kieda**[7], **R. Mukherjee**[5], **D. Nieto**[6], **A. Okumura**[8], **A Petrashyk**[6*], **V. Vassiliev**[1], **for the CTA Consortium**[†]

[1] *University of California Los Angeles*
[2] *Argonne National Laboratory*
[3] *SLAC National Accelerator Laboratory*
[4] *University of Alabama in Huntsville*
[5] *Barnard College, Columbia University*
[6] *Columbia University*
[7] *University of Utah*
[8] *Nagoya University*
*E-mail:* rousselle@astro.ucla.edu

We present the design and the status of procurement of the optical system of the prototype Schwarzschild-Couder telescope (pSCT), for which construction is scheduled to begin in fall at the Fred Lawrence Whipple Observatory in southern Arizona, USA. The Schwarzschild-Couder telescope is a candidate for the medium-sized telescopes of the Cherenkov Telescope Array, which utilizes imaging atmospheric Cherenkov techniques to observe gamma rays in the energy range of 60Gev-60TeV. The pSCT novel aplanatic optical system is made of two segmented aspheric mirrors. The primary mirror has 48 ($\sim$1 m$^2$) mirror panels with an aperture of 9.6 m, while the secondary, made of 24 panels, has an diameter of 5.4 m. The resulting point spread function (PSF) is required to be better than 4 arcmin within a field of view of 6.4 degrees (80% of the field of view), which corresponds to a physical size of 6.4 mm on the focal plane. This goal represents a challenge for the inexpensive fabrication of aspheric mirror panels and for the precise alignment of the optical system as well as for the rigidity of the optical support structure. In this submission we introduce the design of the Schwarzschild-Couder optical system and describe the solutions adopted for the manufacturing of the mirror panels and their integration with the optical support structure.

[*]Presenter

[†]Full consortium author list at http://cta-observatory.org.

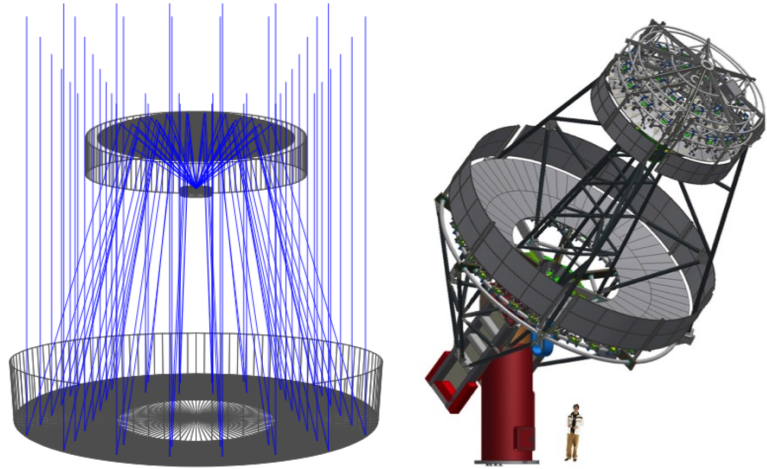

**Figure 1:** Left: Ray-tracing simulation of the Schwarzschild-Couder telescope optical system, which includes the primary and secondary mirrors as well as their respective baffles and the focal plane. Right: CAD model of the full size prototype Schwarzschild-Couder telescope (pSCT) under construction at the Fred Lawrence Whipple Observatory in Arizona.

## 1. Introduction

During the last decade, the scientific achievements of the H.E.S.S., MAGIC and VERITAS observatories have proved the broad scientific value of Imaging Atmospheric Cherenkov Telescopes (IACTs) for the very high energy (VHE) astronomy. The scientific community is now developing the Cherenkov Telescope Array (CTA) [1], the next generation large IACT array made of several tens of telescopes with different apertures. The Schwarzschild-Couder medium-sized telescope (SC-MST) is a candidate telescope for CTA, which utilizes a novel two mirrors optical design with an aperture of 9.6 m. This design offers a wide field of view of 8 degrees, improves the angular resolution compared to current single mirror IACTs and reduces the plate scale of the camera enabling the use of novel SiPM photosensors. However, to achieve these performance improvements it requires a more complex optical system with tighter alignment tolerances than current IACTs [4]. A full size prototype of the SCT-MST (pSCT) is currently under construction at the Fred Lawrence Whipple Observatory in Arizona to test the feasibility of such design as IACT (Figure 1, right).

## 2. Overview of the optical system

The current IACTs, based on Davies-Cotton or parabolic optical systems, have proven to be very efficient and reliable, however these prime focus optical systems also suffer from comatic aberrations when a large field of view is desired. These aberrations can be reduced only by a long focal length resulting in a large camera plate scale and expensive assembly of photodetectors. In comparison, the aplanatic optical system of the Schwarzschild-Couder telescope is not affected by spherical or comatic aberrations [2], and its secondary mirror de-magnifies the image allowing a wide field of view and reduced plate scale (Figure 1, left).

*Schwarzschild-Couder Telescope, Optical System*

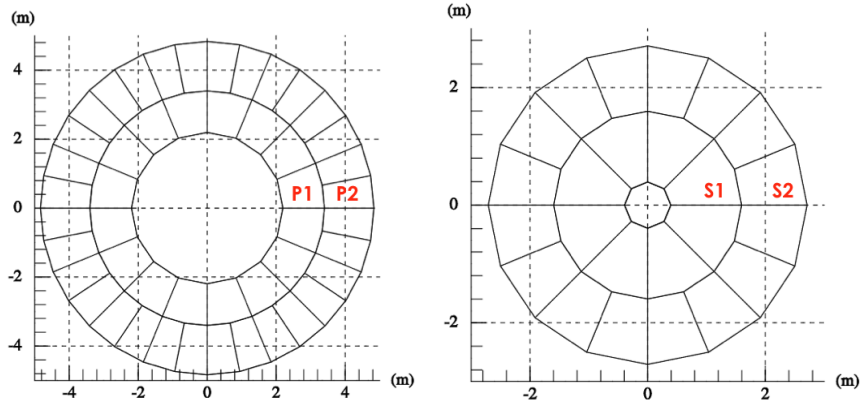

**Figure 2:** The left and right images represent respectively the segmentation schemes of the primary (48 panels) and secondary mirrors (24 panels). Both mirrors use two rings with panels areas between 0.9 and 1.3 m$^2$ (see Table 1).

The definition of the pSCT optical system is presented in Table 1 (left). Its dimensions have been designed to simultaneously provide a large 9.6 m aperture, a field of view of 8 degrees and a PSF smaller than 4 arcmin to match the SiPM pixel size of 6.5 mm in the camera [3].

Both primary (4.83 m radius) and secondary (2.71 m radius) are too large to be inexpensively produced as monolithic mirrors and have to be segmented for cost reduction. The selected segmentation schemes are presented in Figure 2, with mirrors respectively made of 48 and 24 mirror panels of two types. The resulting dimensions of the four different panels types are shown in Table 1 (right). The segmentation of the mirrors is the result of a trade-off between the cost of the complex alignment system and the fabrication cost of large mirror panels. In addition, a gap of 14 mm is introduced between the panels in order to simplify the initial assembly and alignment of the panels on the optical support structure (OSS).

| Focal length [m] | 5.59 |
|---|---|
| Aperture [m] | 9.66 |
| f-number, $f/\#$ | $f/0.57$ |
| Primary radius max [m] | 4.83 |
| Primary radius min [m] | 2.19 |
| Secondary radius max [m] | 2.71 |
| Secondary radius min [m] | 0.39 |
| Field of view [deg] | 8 |

| Mirror panel | P1 | P2 | S1 | S2 |
|---|---|---|---|---|
| Number of panels | 16 | 32 | 8 | 16 |
| Radius max [m] | 3.4 | 4.83 | 1.60 | 2.71 |
| Radius min [m] | 2.19 | 3.40 | 0.39 | 1.60 |
| Diagonal [m] | 1.61 | 1.64 | 1.35 | 1.38 |
| Panel aera [m$^2$] | 1.33 | 1.16 | 0.94 | 0.94 |

**Table 1:** Left: Definition of the pSCT optical system, designed to provide 8 degrees field of view. Right: Definition of the mirror panels. P1 and P2 are respectively the inner and outer mirror panels of the primary mirror, while S1 and S2 correspond to the inner and outer mirror panels of the secondary mirror.

## 3. Primary mirror panels

The cold glass slumping technology, developed by Media Lario Technology (MLT), Italy, has been adapted for the production of the aspheric primary mirror panels [5]. During this process, two 1.7 mm sheets of glass are assembled on each side of a 30 mm aluminum honeycomb core and slumped to required figure on a precisely machined mandrel. The edges of the glass sheets are cut along straight lines before slumping, while still flat, to minimize the stress on the glass.

In 2013 MLT demonstrated the technical feasibility of the slumping process with the successful production of a panel similar to P1 (Table 1), with a 13 mm sag and approximate area of 1 $m^2$ [4]. Four P1 demonstration panels were then produced by MLT, and coated by BTE in Germany. Aluminum was deposited in a vacuum chamber and a barrier layer was used to protect it from oxidation. The mirror panel is then over-coated with a multilayer structure based on silicon and hafnium oxides to increase its longevity and improve its reflectance around 380 nm. For two of these panels, the glass was coated by BTE before the assembly, while the two others were coated after assembly and slumping. This strategy allowed to test two different production chains and identify potential risks, as scratches on the coated glass during slumping, or the time needed for the honeycomb core to reach vacuum during the coating process.

A coordinate measuring machine (CMM), located at the Very High Energy Laboratory at UCLA, was used to calculate the residual error on the mirror figure and slope error for the four P1 panels produced by MLT / BTE. The results show a slope error below 280 $\mu rad$ (over 55 mm averaging scale), which is close but insufficient to meet specifications of <200 $\mu rad$. A deficiency of the P1 mandrel figure has been identified by MLT as the probable main source of error. This mandrel has been resurfaced and its metrology shows a slope error of ∼60 $\mu rad$. Because of its large impact on the pSCT optical performance [4], we anticipate to improve the cold glass slumping process utilizing resurfaced P1 mandrel to bring the slope error of the mirror panels close to 100 $\mu rad$ (goal).

The multilayer coatings produced by BTE on the four P1 panels were measured using a Minolta reflectometer. The absolute reflectance met the specifications at 96% at 380 nm over the full area of the panels, with non-uniformity below 5%.

In addition to the P1 mandrel being resurfaced, the P2 mandrel is now in production and the first panels should arrive at UCLA in November 2015 to be integrated with the alignment system and calibrated.

## 4. Secondary mirror panels

The aspheric secondary mirror of pSCT is significantly curved so that the sag of S1 and S2 panels (Table 1, right) is in the 2-3 cm range. Cold glass slumping technology alone cannot be used to produce such panels and alternative technologies have been investigated.

Four demonstration S2 panels were produced by Flabeg in Germany using hot glass slumping with 12 mm thick glass substrate. Flabeg performed the metrology measurements on these panels using a CMM, and found residuals up to 1 mm (peak to valley) and slope errors of 1.5 mrad, larger than the specifications of 0.2 mm and 0.3 mrad. A part of this error can be attributed to a defect in the bending tool, which was identified by Flabeg, but it is not clear yet if a correction of the tool will be sufficient to meet the specifications.

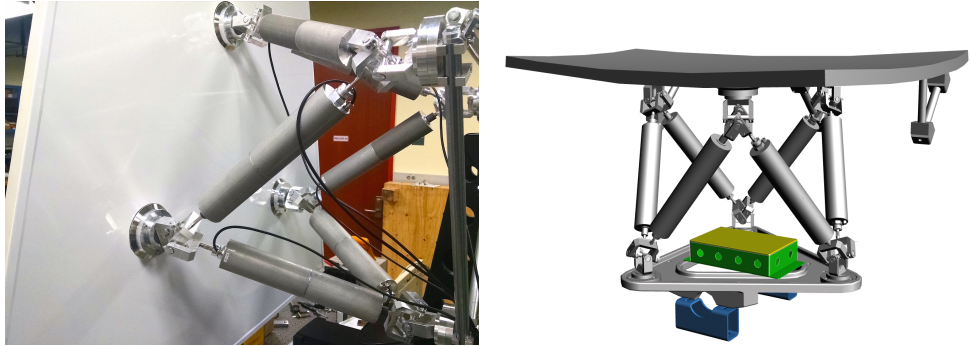

**Figure 3:** Left: Stewart platform holding a P1 panel. Right: CAD model of the final assembly of a mirror panel module, including mirror panel, Stewart platform, controller board, mounting triangle and one edge sensor (far right).

The figure variations observed in four S2 demonstration panels indicate that hot glass slumping alone maybe insufficient to meet the slop error specifications. These considerations motivated an second approach combining hot glass slumping and cold glass slumping to improve the mirror figure. In this process, hot glass slumping is used to pre-shape 2 mm and 12 mm substrates, which are glued together on figuring mandrel during cold glass slumping. This process is currently under development to evaluate its capability to improve the slope error.

## 5. Mirror panel module

The mirror panels of both the primary and secondary mirror are integrated into mirror panel modules (Figure 3, right), which hold and align independently each panel. This module is composed of a Stewart platform (Figure 3, left), a controller board, multiple edge sensors attached on the back of the panel and a mounting triangle (Figure 3, right). A detailed description of the alignment systems and the mirror panel modules can be found in a dedicated proceeding [6].

The controller boards have been assembled and are currently being tested and calibrated, while the assembly of the Stewart platform and edge sensors are still in progress. All the elements of the mirror panel modules are expected to be ready for assembly when the first mirror panels arrive at UCLA in the fall of 2015.

## 6. Assembly on the Optical Support Structure

The different parts of the optical support structure will be assembled first on the ground next to the telescope foundations before being lifted by a crane and installed on the telescope positioner (Figure 4, right) [7]. The primary mirror panel modules will then be installed on the optical support structure while the telescope is pointing at the zenith (primary mirror horizontal). The modules will be lifted using a specially designed fixture (Figure 4, upper left) and installed on brackets welded to the optical support structure, as seen Figure 4 (right). This initial positioning of the mirror panel modules only requires a precision of 10 mm and 1 degree, as the Stewart platforms will correct the remaining misalignment.

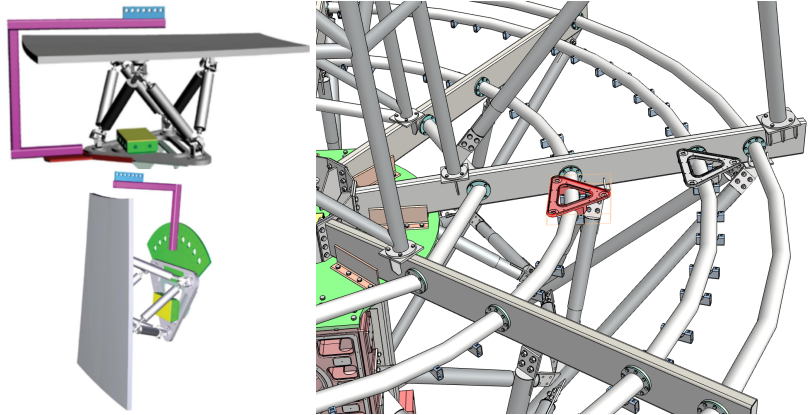

**Figure 4:** Left: The fixtures attached to the primary (top) and secondary (bottom) mirror panel modules are used to lift and install them on the optical support structure. Right: Section of the primary mirror optical support structure. Highlighted in red is one mounting triangle which will support a mirror panel module.

The secondary mirror panel modules will be installed in a similar fashion but with the telescope pointing horizontally and using several dedicated fixtures (Figure 4, lower left) depending on the position of the module around the optical axis. These fixtures, which will have to lift approximately 130 pounds, have been designed and successfully tested using mockup modules.

## 7. Sunlight and stray-light control

To reduce the cost of installation of multiple telescopes in CTA, pSCT will not have a dome to protect it from the sun and weather. The current IACTs mitigate this lack of protection by pointing the telescopes north (in Northern hemisphere) and around 0 degree elevation, avoiding any hazardous reflection of the sun from their single mirror. But the dual-mirror optical system of pSCT cannot use the same strategy due to sun reflections coming from its secondary mirror.

Extensive ray-tracing simulations were used to find a new daylight parking strategy with the objective to avoid any reflected sunlight leaving the optical system. As a result of this study, three parking positions were selected along the year as presented in Figure 5 and two baffles were designed around the primary and secondary mirror, measuring respectively 1.5 and 1 m (Figure 1, right). The addition of three elevation angles for parking ($-5$, 20, 45 deg) and the two baffles provide a safe and affordable solution to avoid injury and fire hazard all year long. We only expect a small amount of reflected sunlight to escape the optical system during few days around the summer solstice, approximately 30 min before sunset. It has been demonstrated through detailed simulations that the light concentration during these few periods remains well below the safety standards.

The baffles are designed to block the sunlight before but also after reflection on the mirrors. As a result, some sunlight will be concentrated inside the baffles with a light concentration smaller than three times the maximum direct sunlight on the ground ($<3500$ W/m$^2$). The baffles are made of sand-blasted aluminum to absorb and diffuse this partially concentrated light, while the area enclosed by the baffles will be restricted for safety reasons.

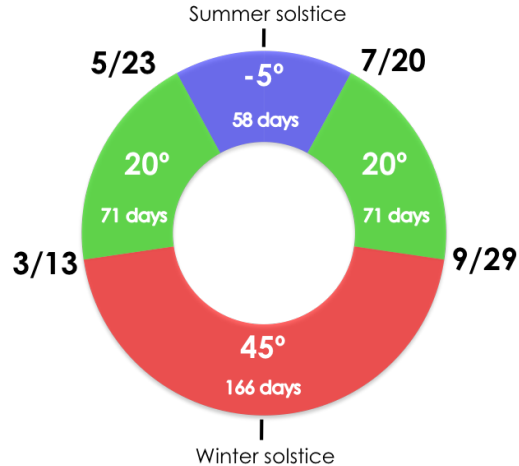

**Figure 5:** Definitions of the three parking positions of pSCT during the year, developed for its construction site in Arizona. The start and end dates are approximate, as the transitions can be made safely in a period of two weeks around these dates. The positions are symmetric around the summer and winter solstices, therefore the spring and fall positions are identical.

In addition to providing safe parking positions, the baffles also reduce the amount of stray light by 98% during the telescope operation, while introducing only 3% of shadowing in the optical system. To reach this small amount of shadowing, the position of the secondary baffle has been optimized and a gap of 30 cm from the edge of the mirror has been introduced, as seen in Figure 1 (left).

## 8. Summary

The prototype Schwarzschild-Couder telescope is now entering the phase of telescope construction. The work on the foundation started in early June 2015 and is expected to be completed in August. The pre-construction activities for pSCT optical system made excellent progress with the production of the demonstration mirror panels, the ongoing assembly of the mirror panel modules as well as the start of the procurement for the primary mirror panels. Special tools have been designed and tested for the installation of the optical system on the optical support structure, which is planned to start in fall 2015.

The technical cost and schedule information learned during the development and implementation of the pSCT project will be critical for the construction of the Schwarzschild-Couder medium-sized telescopes for CTA.

## 9. Acknowledgements

We gratefully acknowledge support from the agencies and organizations listed under Funding

Agencies at this website: http://www.cta-observatory.org/. The development of the prototype SCT has been made possible by funding provided through the NSF-MRI program.